\documentclass[a4paper,12pt]{article}
\usepackage{fancyhdr}
\usepackage{graphicx}
\usepackage{mathptmx}		

\usepackage[affil-it]{authblk} 
\usepackage{etoolbox}
\usepackage{lmodern}
\usepackage{geometry}
\usepackage[round, sort&compress, numbers,super,comma]{natbib} 
\usepackage{ifthen}
\usepackage{url}
\bibpunct{[}{]}{,}{s}{,}{,}  
\title{Influence of the interfacial tension on the microstructural and mechanical properties of microgels at fluid interfaces}

\date{ }

\begin{document}


\begin{center}
\LARGE
\textbf{Influence of the interfacial tension on the microstructural and mechanical properties of microgels at fluid interfaces}
\end{center}

\begin{center}
{\fontsize{12}{14.4}\selectfont
Jacopo Vialetto,$^{a,}$* Natalie Nussbaum,$^{b}$ Jotam Bergfreund,$^{b}$ Peter Fischer,$^{b}$ Lucio Isa$^{a,}$*}
\end{center}

\begin{center}
{\fontsize{11}{13.2}\selectfont
\textit{$^{a}$ Laboratory for Soft Materials and Interfaces, Department of Materials, ETH Z{\"u}rich, Vladimir-Prelog-Weg 5, 8093 Z{\"u}rich, Switzerland.}}

{\fontsize{11}{13.2}\selectfont
\textit{$^{b}$ Institute of Food, Nutrition and Health, ETH Zürich, Schmelzbergstrasse 7, 8092 Z{\"u}rich, Switzerland.}}
\end{center}

\begin{center}
{\fontsize{11}{13.2}\selectfont
* Corresponding authors: jacopo.vialetto@mat.ethz.ch (JV), lucio.isa@mat.ethz.ch (LI)}
\end{center}


\begin{abstract}
Microgels are soft colloidal particles constituted by cross-linked polymer networks with a high potential for applications. In particular, after adsorption at a fluid interface, interfacial tension provides two-dimensional (2D) confinement for microgel monolayers and drives the reconfiguration of the particles, enabling their deployment in foam and emulsion stabilization and in surface patterning for lithography, sensing and optical materials. However, most studies focus on systems of fluids with a high interfacial tension, \textit{e.g.} alkanes/ or air/water interfaces, which imparts similar properties to the assembled monolayers. Here, instead, we compare two organic fluid phases, hexane and methyl tert-butyl ether, which have markedly different interfacial tension ($\gamma$) values with water and thus tune the elasticity and deformation of adsorbed microgels. We rationalize how $\gamma$ controls the single-particle morphology, which consequently modulates the structural and mechanical response of the monolayers at varying interfacial compression. Specifically, when $\gamma$ is low, the microgels are less deformed within the interface plane and their polymer networks can rearrange more easily upon lateral compression, leading to softer monolayers. Selecting interfaces with different surface energy offers an additional control to customize the 2D assembly of soft particles, from the fine-tuning of particle size and interparticle spacing to the tailoring of mechanical properties.
\end{abstract}

\bigskip




\section*{Introduction}

Soft deformable particles are a fascinating class of materials for fundamental~\cite{Yunker2014} and applied research~\cite{Karg2019,Murray2019} alike. Among those, microgels are model soft particles consisting of an internally cross-linked polymer network that swells in a good solvent. With respect to hard, mechanically rigid particles (\textit{e.g.}, polystyrene or silica colloids), they offer intriguing advantages stemming from their ability to deform and reconfigure in response to a variety of stimuli.~\cite{Geisel2014,Peng2015,Gnan2019}
These properties are often retained at fluid interfaces, where microgels have been extensively studied in relation to multiple applications, such as the preparation of responsive foams~\cite{Fujii2011} and emulsions~\cite{Richtering2012} as well as the surface patterning of ordered structures for lithography,~\cite{Fernandez-Rodriguez2021} sensing~\cite{Kim2005} or optics.~\cite{Tsuji2005}
In particular, as opposed to rigid particles, the deformability of microgels enables the robust fabrication of two-dimensional (2D) non-close-packed hexagonal arrays of features with continuously varying spacing~\cite{Rey2016nano} and the realization of complex, non-triangular patterns by tailoring the softness of the interaction potential, both from single~\cite{Rey2017,Fernandez-Rodriguez2021} and sequential depositions.~\cite{Grillo2020}

The most widely studied microgels are made of poly-N-isopropylacrylamide (pNIPAM) prepared by precipitation polymerization. When swollen in a good solvent, such as water below the volume phase transition temperature, they comprise a more cross-linked, denser core, a less cross-linked shell, and uncross-linked chain ends and loops forming an external "fuzzy" surface.~\cite{Stieger2004,Bergmann2018,Ninarello2019} After adsorption onto a fluid interface, the outer polymer chains flatten on the interface plane to minimize the interactions between the fluids, while the more crosslinked core remains mostly in the water phase, and the particle assume a "core-corona" morphology.~\cite{Geisel2012,Camerin2019} 
Recent results show that an improved control over the final assemblies is provided by a rational control of the particle morphology as obtained from synthesis.~\cite{Ciarella2021} This allows, for example, to control structural phase transitions in microgel monolayers upon compression depending on the crosslinking density of the internal core.~\cite{Vialetto2021} 

However, up to now, most studies addressing microgels' conformation, assembly and interfacial properties focused on microgels adsorbed either at air/water or alkane/water interfaces.~\cite{Harrer2019,Bochenek2021} While the effect of different fluids has been studied in great details for hard colloids,~\cite{Binks2000,Bergfreund2019,Vialetto2021FA} and reconfigurable objects such as proteins,~\cite{Bergfreund2021,Bergfreund2021SM} it has been rarely taken into account for microgels.~\cite{Schmidt2011,Destribats2011,Rumyantsev2016}
In particular, air and alkane interfaces with water share common properties that influence the behavior of adsorbed microgels. Both types of fluids are non-solvents for pNIPAM and consequently the polymers in contact with the top phase are in a fully collapsed state. Additionally, both interfaces are characterized by relatively high values of interfacial tension ($\gamma$), 72 and $\simeq 50 mN \cdot m^{-1}$ for air and alkane/water interfaces, respectively. This imparts similar conformations to the microgels upon adsorption, and, consequently, an  analogous structural and mechanical behavior of adsorbed monolayers upon interfacial compression.~\cite{Harrer2019,Bochenek2021} 
Here, we show that by using different organic phases, the fluid interface itself can be engineered to control the shape of adsorbed microgels and the resulting properties of interfacial assemblies. As the organic phase, we choose solvents with markedly different interfacial tension values to water, focusing on hexane and methyl tert-butyl ether (MTBE), having $\gamma$ = 50.4 and 9.8 $mN \cdot m^{-1}$, respectively. 
On the other hand, given their similar Hildebrand solubility parameters,~\cite{Hansen2007} we expect both of them to be non-solvents for pNIPAM,~\cite{Yagi1992}, therefore allowing us to decouple the effect of a marked surface tension variation, with that of a change in solubility of the polymer network. By using a combination of atomic force microscopy (AFM) and controlled monolayer compressions in a Langmuir–Blodgett trough, we analyze the microstructural conformation of individual microgels as a function of the fluid interface, and relate it to the resulting structural features and mechanical properties of the monolayers. These investigations identify additional factors to control the assembly of microgels at fluid interfaces in view of the aforementioned applications, but also offer insights into the complex interactions of soft objects at fluid interfaces, of potential interest for a range of phenomena where soft colloids (including biological particles) are involved. 

\section*{Materials and Methods}
\small
\subsection*{Reagents}

N,N'-Methylenebis(acrylamide) (BIS, Fluka 99.0\%), methacrylic acid (MAA, Acros Organics 99.5\%), potassium persulfate (KPS, Sigma-Aldrich 99.0\%), isopropanol (Fisher Chemical, 99.97\%), toluene (Fluka Analytical, 99.7\%), n-hexane (SigmaAldrich, HPLC grade 95\%) and methyl tert-butyl ether (MTBE, SigmaAldrich, ACS reagent > 99.5\%) were used without further purification. 
N-isopropylacrylamide (NIPAM, TCI 98.0\%) was purified by recrystallization in 60/40 v/v toluene/hexane. 

\subsection*{Microgel synthesis}
The microgels used in this study were synthesized by free-radical precipitation polymerization. 

\noindent \textit{Soft microgels.} NIPAM (0.385 g), 5 mol \% MAA and 1 mol \% BIS were dissolved in 25 mL of MQ water at room temperature. The reaction mixture was then immersed into an oil bath at 80 $^{\circ}$C and purged with nitrogen for 1 h. The reaction was started by adding 10 mg of KPS previously dissolved in 1 mL MQ water and purged with nitrogen. The polymerization was carried out for 6 h in a sealed flask. Afterwards, the colloidal suspension was cleaned by dialysis for a week, and 8 centrifugation cycles and resuspension of the sedimented particles in pure water. 

\noindent \textit{Stiff microgels.} NIPAM (1 g), 5 mol \% MAA and 5 mol \% BIS were dissolved in 50 mL of MQ water at room temperature. The reaction mixture was then purged with nitrogen for 1 h. Afterwards, 40 mL of the monomer solution was taken out with a syringe. 10 mL of MQ water were added to the reaction flask and the solution was immersed into an oil bath at 80 $^{\circ}$C and purged with nitrogen for another 30 min. The reaction was started by adding 13 mg of KPS previously dissolved in 1 mL MQ water and purged with nitrogen. After 1.5 minutes the solution turned slightly milky, and the feeding of the monomer solution (40 mL at 1.5 $mL \cdot min^{-1}$) to the reaction flask was started. When the feeding was terminated, the reaction was immediately quenched by opening the flask to let the air in, and placing it in an ice bath. The obtained colloidal suspension was cleaned by dialysis for a week, and by 8 centrifugation cycles and resuspension of the sedimented particles in pure water.

\subsection*{Methods}

\noindent \textit{DLS and SLS.} Dynamic light scattering (DLS) experiments were performed using a Zetasizer (Malvern, UK). The samples were let to equilibrate for 15 min at the required temperature (22 or 40$^{\circ}$C) prior to performing six consecutive measurements. For static light scattering (SLS), a CGS-3 Compact Goniometer (ALV, Germany) system was used, equipped with a Nd-YAG laser, $\lambda = 532$ nm, output power 50 mW before optical isolator, measuring angles from 30$^{\circ}$ to 150$^{\circ}$ in 2$^{\circ}$ steps. Static scattering form factor analysis was performed using the FitIt! tool developed by Otto Virtanen for MATLAB.~\cite{Virtanen2016} A detailed description of the fitting procedure is reported elsewhere.~\cite{Vialetto2021}

\noindent \textit{Deposition of isolated microgels from the fluid interface.} Microgels were deposited from the fluid interface onto silicon wafers for atomic force microscopy (AFM) imaging following an already reported procedure.~\cite{Vialetto2021}
Silicon wafers were cut into pieces and cleaned by 15 min ultrasonication in toluene, isopropanol, acetone, ethanol and MQ water. A piece of silicon wafer was placed inside a Teflon beaker on the arm of a linear motion driver and immersed in water. Successively, a liquid interface was created between MQ water and n-hexane or MTBE. Around 10 $\mu$L of the microgels suspension was injected at the interface after appropriate dilution in a 4:1 MQ-water:IPA solution. After 10 min equilibration time, extraction of the substrate was conducted at a speed of 25 $\mu m \cdot s^{-1}$ to collect the microgels by crossing the fluid interface.

\noindent \textit{Langmuir trough deposition.} Microgels assembled at the fluid interface at controlled surface pressure ($\Pi$) values were deposited onto silicon wafers for visualization using a custom-made setup already reported in literature~\cite{Rey2016}. We used a KSV5000 Langmuir trough equipped with a dipper arm immersed in water for holding a silicon substrate forming an angle of approximately 30$^{\circ}$ with the water surface. The silicon substrate was further cleaned in a UV-Ozone cleaner (UV/Ozone Procleaner Plus, Bioforce Nanosciences) for 15 min to ensure a hydrophilic surface prior to microgel deposition. After forming an interface between water and hexane or MTBE, the substrate was lifted so to pierce the liquid interface. Microgels were then injected on the liquid interface while the surface pressure was simultaneously measured with a platinum Wilhelmy plate. When the required initial surface pressure was reached, the injection was stopped and the interface was left to equilibrate for 15 min. Successively, the dipper was activated to extract the substrate at a constant speed of 0.3 $mm \cdot min^{-1}$ and, after 2 min, the barriers started moving at a compression speed of 2.3 $mm \cdot min^{-1}$. When the compression finished, the barriers were immediately opened while the substrate was still moving up in order to achieve a discontinuity in microgel concentration deposited on the silicon wafer. Due to the finite size of the compressible area in the trough, multiple experiments were required to obtain the full compression isotherms.

The conformation of microgels at the interface and their 2D assembly as a function of the surface pressure was then inferred by analysing the substrates using atomic force microscopy (AFM). Images from the initial position of the three-phase contact line to the end of the substrate were recorded at a fixed distance of 500 $\mu$m. The discontinuity in microgels deposition ensures a correct assignment of the surface pressure value measured at the liquid interface during compression to the corresponding position on the silicon substrate. More specifically, the highest value of surface pressure measured during the experiment was assigned to the position on the substrate corresponding to the highest density of microgels. Consequently, knowing the dipper speed and the distance between AFM images of the substrate, the surface pressure curve was scanned backwards assigning to each AFM image its corresponding value of $\Pi$.

\noindent \textit{AFM imaging and analysis.} Microgels deposited on silicon wafers were characterized by AFM (Bruker Icon Dimension), in tapping mode, using cantilevers with 300 kHz resonance frequency and 26 $mN \cdot m^{-1}$ spring constant. Height and phase images were recorded at the same time. Images were first processed with Gwyddion and successively analysed with custom MATLAB codes. 
The following procedure was used to obtain the averaged microgel height profile. For each microgel, horizontal and vertical profiles passing through its center were extracted from AFM height images. Successively, an average over around 20 microgels was obtained by aligning each profile by its center value. The diameter of the microgels at the interface ($\sigma$) was calculated by fitting the phase images with a circle. The core diameter of individual microgels was measured setting a lower threshold in the height profiles at 2.5 nm. From this value we calculated the lateral extension of the corona ($\delta$).

For the height profiles of single microgels in 2D assemblies as function of $\Pi$, the same procedure was used; the profiles were then cut on the r-axis to exclude neighbouring microgels. The core size of the microgels in compressed monolayers ($\sigma_{core}$) was estimated by setting a lower threshold in the height profiles at approximately 15 nm.

The average inter-particle distance $d_{cc}$ at different $\Pi$ was estimated by extracting the positions of the microgels from AFM images taken at different locations on the substrates. For a given set of particles' coordinates (x,y), $d_{cc}$ was calculated as the average distance between neighbouring particles. The neighbours' list was constructed based on the Voronoi tessellation using the Freud open-source Python libraries~\cite{eric_s_harper_2016_166564}.
Across the isostructural solid-solid phase transition, two populations of particles with distinct interparticle distances are evidenced; namely, particles in core-core contacts, and particles separated by their polymeric coronae. A threshold value was used to separate the two populations. 
Such neighbours' list was also used to calculate the average hexatic order parameter parameter $\Psi_6$:
\begin{equation}
  \Psi_6=\left\langle \frac{1}{N_j}\sum_{k=1}^{N_j} e^{i6\theta_{jk}}\right\rangle
\end{equation}
$N_j$ is the number of neighbours of the j-th particle in the AFM image, $\theta_{jk}$ is the angle between the unit vector (1,0) and the vector $\textbf{r}=\textbf{r}_k-\textbf{r}_j$ connecting particle $j$ and its k-th neighbour.

\normalsize

\section*{Results and Discussion}

We investigate the microstructural and mechanical properties of microgel assemblies at different fluid interfaces by assembling monolayers at flat interfaces in a Langmuir–Blodgett trough, where the monolayer compression and structural organization can be precisely monitored. During compression, we transfer the microgel assemblies onto a silicon wafer that is lifted through the fluid interface forming a 30$^\circ$ angle with respect to the interface plane.~\cite{Rey2016} This simultaneous compression and deposition allow us to continuously vary and monitor both the surface pressure ($\Pi$) as a function of the trough area and the resulting particle assembly at different packing fractions. Information of the monolayer microstructure is then obtained by imaging the dried substrates by AFM and, by knowing the position on the substrate and the lifting speed, we can relate each AFM image with its $\Pi$ value. Unless otherwise stated, the microgels used in this study are made with 1 mol \% of N,N'-Methylenebis(acrylamide) (BIS) as crosslinker. A characterization of their bulk properties by dynamic and static light scattering is reported in Table S1 and Fig. S1.

\begin{figure*}
 \centering
 \includegraphics[scale=0.9]{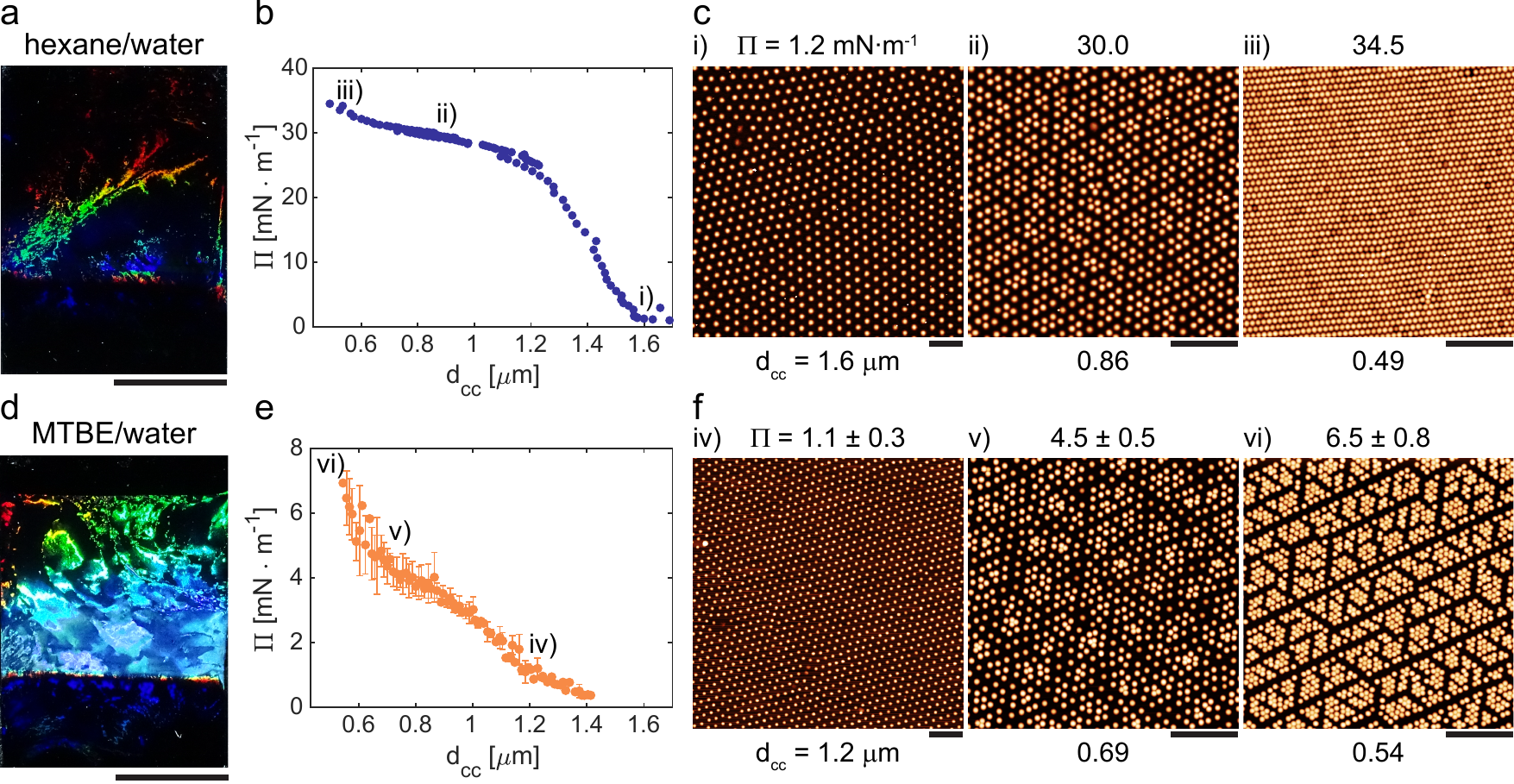}
 \caption{Two-dimensional assembly of microgels as function of interfacial compression, for different oil/water interfaces. a) Photograph of a silicon substrate after deposition of a microgel monolayer from the hexane/water interface. Scale bar: 1 cm. b) Compression curve at the hexane/water interface as function of the average interparticle distance ($d_{cc}$). c) AFM height images of the monolayer transferred from the hexane/water interface onto a silicon wafer, at increasing surface pressure ($\Pi$), and corresponding decrease of $d_{cc}$. d) Photograph of a silicon substrate after deposition of a microgel monolayer from the MTBE/water interface. e) Compression curve at the MTBE/water interface as function of $d_{cc}$. f) AFM height images of the monolayer transferred from the MTBE/water interface onto a silicon wafer. All scale bars for AFM images: 5 $\mu$m. }
 \label{Fig.assembly}
\end{figure*}

Two substrates prepared from hexane/water and MTBE/water interfaces, are shown in Fig.~\ref{Fig.assembly}(a, d). The visual appearance of the two samples, imaged under the same illumination conditions and prepared at similar relative compressions, qualitatively reveals that, in both cases, the particles display long-range ordering, as indicated by the presence of structural colors. However, the different brightness of those structural colors implies differences in the interparticle distance, and/or refractive index, between the two microgel assemblies.

We inspect the assembly structure as a function of compression (Fig.~\ref{Fig.assembly}(b, e)). The behavior of $\Pi$ \textit{versus} the average interparticle distance ($d_{cc}$), for monolayers at the hexane/water interface, follows the behavior already reported for other microgels adsorbed at the same fluid interface (Fig.~\ref{Fig.assembly}(b)).~\cite{Rey2016,Rey2017} At low interfacial compression and particle concentration, microgels are dispersed on the fluid interface and form a disordered structure with interparticle distances larger than their size ($d_{cc} > 1.6 \mu m$ and $\Pi \leq 1 mN \cdot m^{-1}$, Fig. S2). Upon increasing compression, all adsorbed microgels enter in contact through their extended coronae and arrange into a long-ranged hexagonal assembly. A further decrease of the trough area causes a steep increase of the surface pressure, corresponding to compression of the monolayer and to a continuous decrease of the lattice constant of the hexagonal assembly. Subsequently, the monolayer undergoes an isostructural solid-solid phase transition, with the nucleation of clusters of particles at much shorter separation distances. This transition is accompanied by a reduction in the slope of the compression curve. Ultimately, a last kink in the measured surface pressure is observed when all particles enter into core-core contacts, forming a close-packed hexagonal assembly at high compression. Examples of the different monolayer structures as a function of $\Pi$ are reported in Fig.~\ref{Fig.assembly}(c).

When microgels assemble at the MTBE/water interface, the whole surface pressure curve (Fig.~\ref{Fig.assembly}(e)) is shifted to lower values due to the much lower value of $\gamma$ for a bare MTBE/water interface. Qualitatively, a similar microstructural phase behavior throughout the compression isotherm is evidenced (Fig.~\ref{Fig.assembly}(f)). At low $\Pi$ values, there is a coexistence of disordered regions with interparticle separations larger than the particle size, together with long-range hexagonally packed zones with microgels in contact through their extended coronae. The hexagonally packed assembly is characterized by lower interparticle distances with respect to the same structure at the hexane/water interface, indicating that the microgel undergoes a less pronounced deformation and is effectively smaller at such interface. Upon increasing compression, all microgels assemble in an 2D hexagonal lattice with decreasing interparticle distance. Successively, as for the monolayers at the hexane/water interface, a further decrease of the trough area induces an isostructural phase transition. However, in our experiments, we could not reach values of $\Pi$ higher than $6-7 mN \cdot m^{-1}$. Therefore, it is unclear if the close-packed assembly found in monolayers at the hexane/water interface is formed for MTBE/water interfaces as well. The absence of a close-packed assembly region at high compression could be attributed to the forced desorption of microgels due to the lower $\gamma$ (and consequently lower desorption energies). Alternatively, it may originate from the size of the compressible area in the trough, which is not enough to fully compress all microgels that could be added to the interface prior to compression.~\cite{Pinaud2014} While the microstructural organization is qualitatively similar for the two oils investigated, a distinct variation of the mechanical response of the monolayer as a function of $d_{cc}$ is evidenced by the different slopes of the compression curves, as we further analyze below.

In particular, we first examine the conformation of individual microgels for the two values of $\gamma$. An indirect measure of the microgels’ profiles at the fluid interface is obtained by keeping the surface pressure low enough to deposit individual, uncompressed particles, and by characterizing their height profiles by AFM in the dried state. 
\begin{figure}
\centering
  \includegraphics[scale=1]{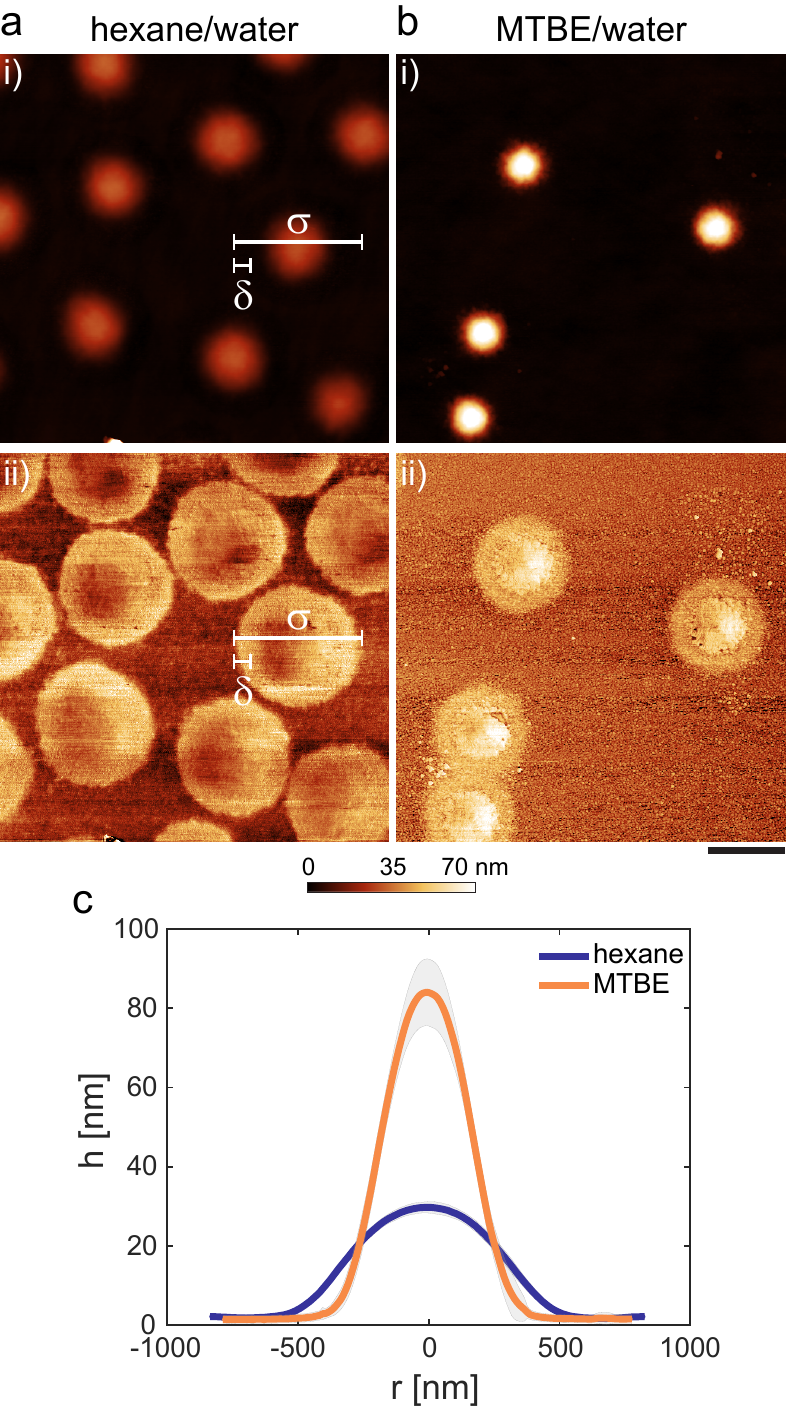}
  \caption{Conformation of individual microgels adsorbed at the oil/water interface and transferred onto a silicon wafer. a-b) Representative AFM height (i) and phase (ii) images of microgels transferred from the hexane/water (a, left column) or MTBE/water (b, right column) interface. Scale bar: 1 $\mu$m. c) Experimental height profiles of microgels transferred from the hexane/water (blue curve) or MTBE/water (orange curve) interface. The shaded regions correspond to the standard deviations of the height profiles calculated on around 20 particles.}
  \label{Fig.single_ug}
\end{figure}
In both cases (Fig.~\ref{Fig.single_ug}(a-b)), the microgels exhibit the typical core-corona profile.\cite{Geisel2012,Camerin2019,Vialetto2021} However, the height profiles are quantitatively different. Specifically, the particles deposited from the hexane/water interface reach a maximum dry height of $h = 30 \pm 1.5 nm$ (Fig.~\ref{Fig.single_ug}(c), blue curve), while the same quantity increases by roughy three times at the MTBE/water interface, reaching $h = 84 \pm 8 nm$, (Fig.~\ref{Fig.single_ug}(c), orange curve). In either case, the height profile smoothly decays towards the particle periphery until it is reduced to a very thin layer comprising the outer uncrosslinked polymer chains, which expand on the interface plane and become visible only in the phase images. The overall lateral extension of the microgels is different for the two fluids. The total lateral extension of the particles adsorbed at the hexane/water interface, as obtained from phase images, is $\sigma = 1.54 \pm 0.05 \mu m$, while for the MTBE case $\sigma$ decreases to $1.26 \pm 0.04 \mu m$. The extent of the in-plane deformation at the fluid interface can be referred in relation to the particle size in bulk aqueous suspensions by computing the ratio $\sigma/D_{h}$, where $D_{h}$ is the hydrodynamic diameter measured with dynamic light scattering (Table S1). This stretching ratio increases from 1.60 $\pm$ 0.08 to 1.96 $\pm$ 0.09 from MTBE/water to hexane/water interface. 

Moreover, by identifying the outer corona as the region of the microgels where the measured height is below 2.5 nm, we can detect the microgel core (considered as where the majority of the polymer composing the particle is) and the lateral extension of the corona ($\delta$). Interestingly, we measure an identical extension of the corona $\delta = 0.22 \pm 0.03 \mu m$ irrespective of the oil we used, while the corresponding core size changes from $1.32 \pm 0.05 \mu m$ for the hexane case to $1.02 \pm 0.04 \mu m$ for the MTBE case.

These data clearly show that the polymer network undergoes a different rearrangement upon using different oil phases. In particular, both the height and the lateral extension of the adsorbed microgels depend on the oil phase, with a higher $\gamma$ value causing an increased in-plane stretching of the particles and corresponding decrease of the maximum height, as predicted by elastocapillary models.~\cite{Style2015}
This indicates that the crosslinked core of the particle has an internal elasticity that counterbalances the deformation imposed by interfacial tension, undergoing different deformations depending on the energy of the interface. Conversely, a constant extension of the external coronae is a direct consequence of the ability of the uncrosslinked chain ends to expand unconstrained onto the interface plane in order to minimize contacts between the two fluids. 
Notably, when using toluene as the top phase ($\gamma$ = 36.3 $mN \cdot m^{-1}$), the resulting dried profiles resemble that obtained at the hexane/water interface (Fig. S3), including $\delta = 0.22 \pm 0.04 \mu m$. This indicates that a value of $\gamma \geq 36 mN \cdot m^{-1}$ is enough to cause maximum stretching of the adsorbed microgels on the interface plane. 
This result is corroborated by earlier works by Camerin et al.~\cite{Camerin2019} and Harrer et al.,~\cite{Harrer2019} which reported similar AFM profiles for microgels adsorbed at the benzene/water, decane/water, or air/water interfaces ($\gamma = 36, 51$ and $72 mN \cdot m^{-1}$, respectively). Instead, it differs from what reported by Bochenek et al.,~\cite{Bochenek2021} which found an increase in the particle in-plane diameter from the decane/water to the air/water interface. As already mentioned by these authors, such variations can be attributed to differences in the bulk particle size, as well as, possibly, in the particle internal morphology.

We also investigated the effect of the internal microgel elasticity on the resulting conformation at the different fluid interfaces, by analyzing dried profiles of stiffer microgels made with 5 mol \% BIS (Fig. S4). As already reported,~\cite{Camerin2019} stiffer microgels adsorbed from the hexane/water interface are thicker with respect to softer ones (reaching $h = 122 \pm 4 nm$). They also maintain a similar core-corona profile, with a smaller corona thickness of $\delta = 55 \pm 33 nm$. In our case, variations in particle conformation when MTBE was used was similar for both softer and stiffer microgels. The maximum height of the latter increased by roughy three times, reaching $h = 320 \pm 9 nm$, while the extension of the corona remained constant ($\delta = 59 \pm 43 nm$). Overall, the exact values of $h$, $\sigma$ and $\delta$ depend on the microgel stiffness, instead, the relative differences with respect to the interfacial tension of the fluid interface are similar.

\begin{figure}
\centering
  \includegraphics[scale=1]{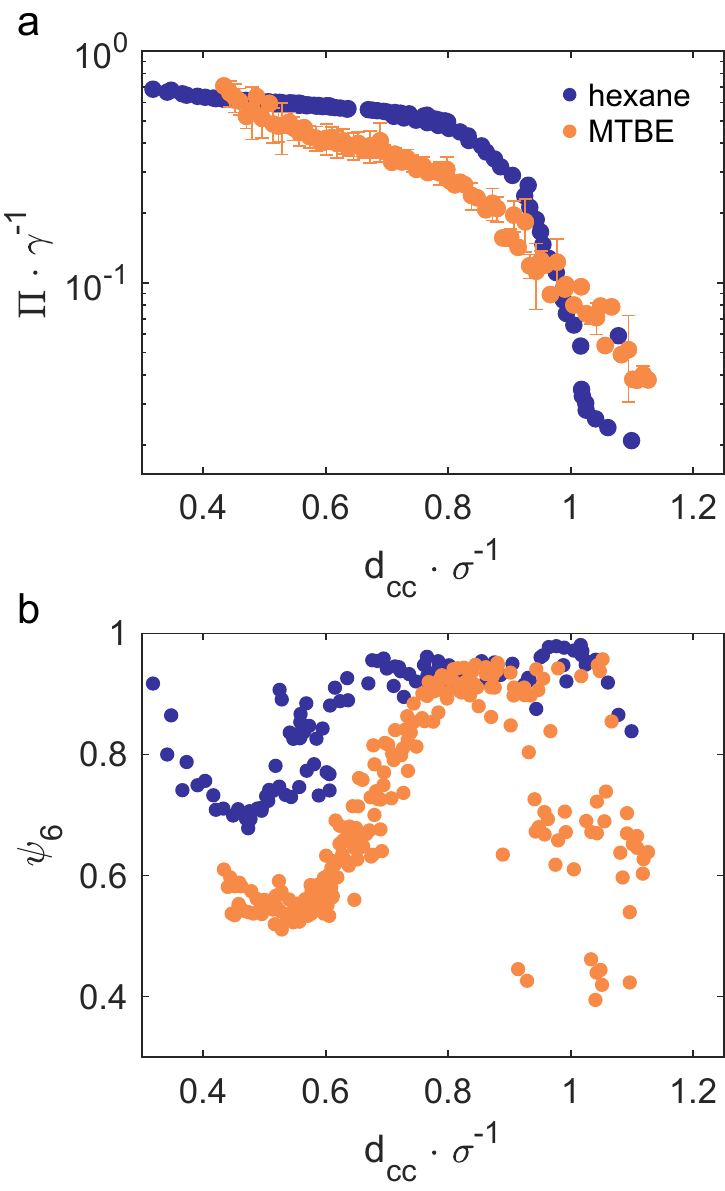}
  \caption{Normalized compression curves and hexatic bond-order parameter ($\Psi_6$) for different oil/water interfaces. a) Surface pressure normalized by the interfacial tension of the bare fluid interface \textit{versus} the interparticle distance normalized by the diameter of an adsorbed microgel prior to compression, at the hexane/water (blue points) or MTBE/water (orange points) interface. b) Average hexatic order parameter $\Psi_6$ as function of the normalized $d_{cc}$.}
  \label{Fig.comp_curve}
\end{figure}

The analysis of individual particles enables us to renormalize the mechanical response of the monolayers, taking into account the single particle in-plane deformation. In Fig.~\ref{Fig.comp_curve}(a) we plot the compression curves normalizing $d_{cc}$ with the respective values of $\sigma$, and $\Pi$ normalized with the $\gamma$ values. The normalized curves exhibit markedly different slopes as a function of the oil phase used. This indicates a variation in the monolayer mechanical properties, which do not depend on the single particle conformation, but on their collective response upon interfacial compression, as connected to its structure (see Fig.~\ref{Fig.assembly}(c) and (f)). At low compression, in the region spanning from individual microgels to 2D crystals ($1.1 < d_{cc} \cdot \sigma^{-1} < 0.8$), the slope of the compression curve at the MTBE/water interface is lower than that at the hexane/water interface. This can be attributed to an increased ability of the microgels to accommodate deformations for lower $\gamma$ values. Consequently, the monolayer is softer and less work is required to compress the 2D hexagonal structure in this region. A closer look at the microgels’ conformation within the monolayer, up to the isostructural phase transition, is reported in Fig. S5. In this regime, all microgels in the assembly compress uniformly within the interface plane. The height profiles as a function of $\Pi$ (Fig. S5) show that, at both fluid interfaces, the in-plane size decreases while the particles' height increases. However, the particles' aspect ratio (Fig. S5(c)), measured as $d_{cc} \cdot h^{-1}$, clearly indicates that, for comparable compression, the microgels at the MTBE/water interface are more deformable within the interface plane, as their aspect ratio show a much weaker dependence on compression than for microgels at the hexane/water interface. The different mechanical response of the microgel monolayers as a function of $\gamma$ is also captured by plotting the normalized surface pressure \textit{versus} the average compressive strain in the monolayers ($\epsilon$), where $\epsilon = (\sigma - d_{cc}) / \sigma$ (Fig. S6). At similar compressive strain, microgels at the hexane/water interface yield stiffer monolayers.

Normalizing $d_{cc}$ with $\sigma$ makes it also possible to compare the various degrees of structural order in the assemblies at equal relative compression. In particular, we calculate the average hexatic order parameter $\Psi_6$ as a function of compression (Fig.~\ref{Fig.comp_curve}(b)). $\Psi_6$  describes the average degree of 6-fold symmetry in the structures, whereas a value of 1 indicates a perfect hexagonal arrangement. Upon normalization, as expected, highly ordered hexagonal lattices are obtained at $d_{cc} \cdot \sigma^{-1} \simeq 1$, i.e. when microgels enter into contact through their extended coronae, at both interfaces. Interestingly, $\Psi_6$ starts to decrease earlier for monolayers at the MTBE/water interface compared to that at the hexane/water interface, indicating that the onset of the isotructural phase transition is at larger relative interparticle distances in the former case. In practice, a lower level of compression is required for partial collapse of some of the microgels’ coronae at the MTBE/water interface. In this case, the phase transition region is broader, and $\Psi_6$ reaches much lower values with respect to the assembly at the hexane/water interface. These traits indicate the existence of an interplay between single-particle conformation and the characteristics of the phase transition (see below).
The phase transition is gradually accompanied by a variation of the slope of the surface pressure curve, which ultimately becomes steeper when MTBE is used as the upper phase (Fig.~\ref{Fig.comp_curve}(a), $d_{cc} \cdot \sigma^{-1} > 0.6$). 
Notably, this upturn in the slope is accompanied by an incipient increase of the $\Psi_6$ values, suggesting the possible completion of the isostructural phase transition at higher compression.

\begin{figure}
 \centering
 \includegraphics[scale=1]{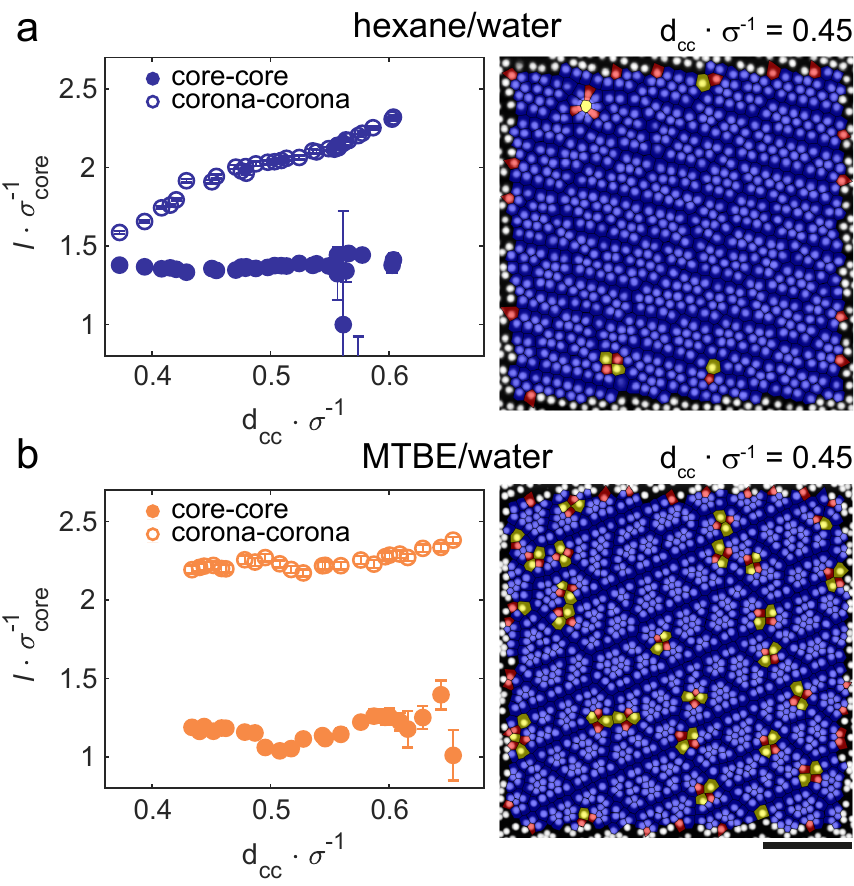}
 \caption{Structural characterization of the microgel assemblies at high interfacial compression.
 a) Left: nearest-neighbor distance $l$, normalized by the size of the particles' cores ($\sigma_{core}$), as a function of the normalized $d_{cc}$, across the isostructural phase transition, for a monolayer at the hexane/water interface. Filled symbols: distance between particles in core-core contact. Open symbols: distance between particles separated by their coronae. Right: representative AFM image of the particle monolayer at the hexane/water interface at $d_{cc} \cdot \sigma^{-1} = 0.45$. In red, blue and yellow particles with 5, 6 and 7 nearest neighbors, respectively. 
 b) Left: normalized nearest-neighbor distance \textit{versus} the normalized $d_{cc}$, across the isostructural phase transition, for a monolayer at the MTBE/water interface. Right: representative AFM image of the particle monolayer at the MTBE/water interface at $d_{cc} \cdot \sigma^{-1} = 0.45$. Scale bar for all images: 5 $\mu$m. }
 \label{Fig.struct_analysis}
\end{figure}


For a more detailed characterization of the structures across the isostructural phase transition, we plotted the nearest-neighbor distance $l$, normalized by the size of the particles' cores ($\sigma_{core}$), as a function of the normalized $d_{cc}$ (Fig.~\ref{Fig.struct_analysis}(a-b)). The population of nearest-neighbor distances shows two separate values throughout the phase transition, corresponding to particles in core-core contacts (filled circles), and particles that are separated by the polymer composing their coronae (open circles).~\cite{Rey2016} Structures at the hexane/water interface display constant core-core separations, while the distance between particles in corona-corona contacts decreases constantly (Fig.~\ref{Fig.struct_analysis}(a)). Conversely, both quantities remain approximately constant at the MTBE/water interface at all $d_{cc} \cdot \sigma^{-1}$ values investigated (Fig.~\ref{Fig.struct_analysis}(b)).
These results can be rationalized considering a different response of the polymer network at high interfacial compression. At the hexane/water interface, failure of the coronae in some directions causes the formation of clusters with particles at shorter separation distances. Part of the polymer composing the coronae remains on the fluid interface separating the cores until the isostructural phase transition is not completed, maintaining the cores at an averaged distance of $l \cdot \sigma_{core}^{-1} = 1.3 \pm 0.2 \mu m$ (filled circles in Fig.~\ref{Fig.struct_analysis}(a)). The stretched corona in the other directions is continuously compressed within the interface plane, decreasing constantly the average separation between clusters. Complete collapse of the coronae happens only at the end of the isostructural phase transition, when all cores are forced to enter into contacts and the monolayer becomes a closely-packed 2D crystal. 
Instead, the corona of microgels at the MTBE/water interface either remains fully stretched, or collapses onto the cores, presumably desorbing from the interface into the water phase. This allows for the formation of clusters of particles in core-core contact, having an averaged distance of $l \cdot \sigma_{core}^{-1} = 1.14 \pm 0.08 \mu m$ (filled circles in Fig.~\ref{Fig.struct_analysis}(b)), the number of which increases upon further interfacial compression, but the distance between such clusters remains approximately constant.
Such a difference in particle organization can be directly visualized in representative AFM images at the same, relative, interparticle separation (Fig.~\ref{Fig.struct_analysis}(a-b)).

\begin{figure}
 \centering
 \includegraphics[scale=1]{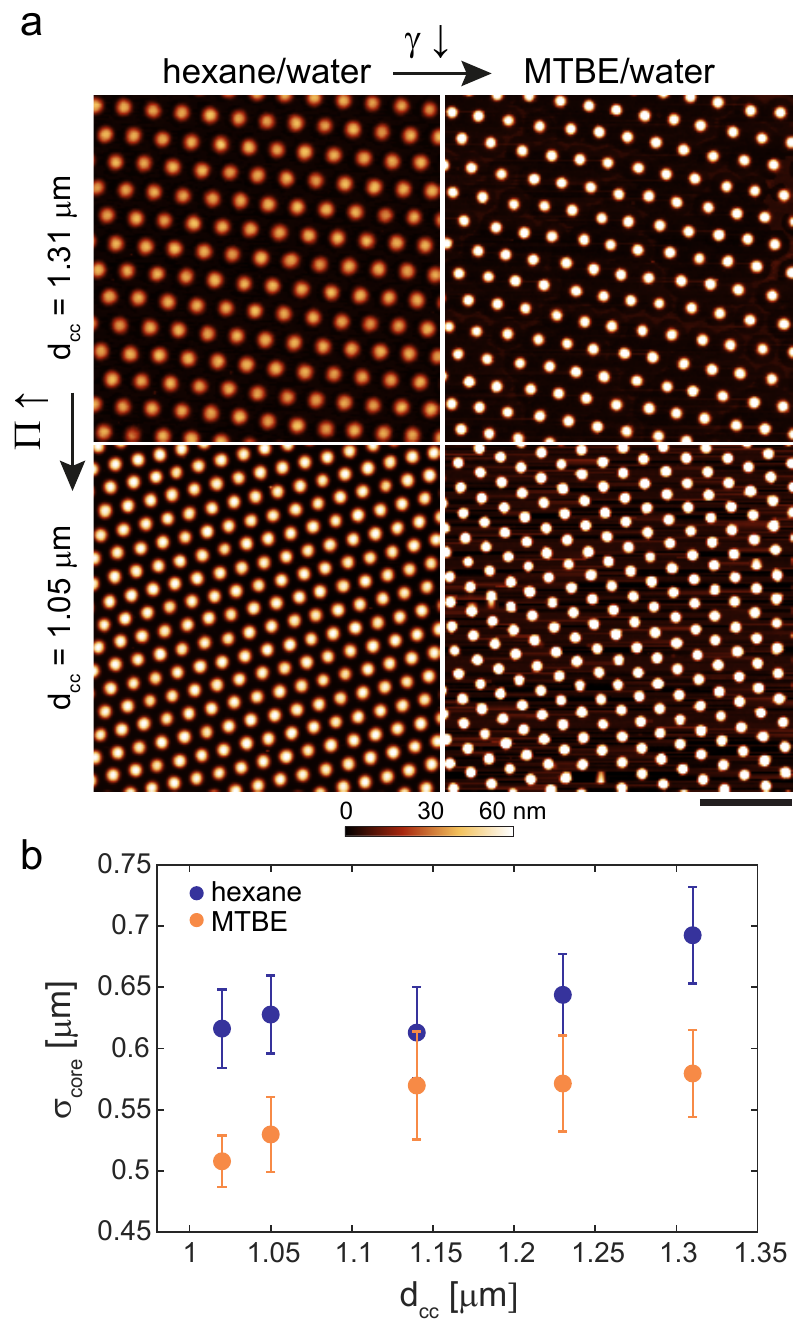}
 \caption{Comparison of the crystalline structures formed at the different oil/water interfaces at low to intermediate monolayer compression. 
 a) Representative AFM images of microgels transferred from the hexane/water (left column) or MTBE/water (right column) interface. Top row: $d_{cc} = 1.31 \mu m$; bottom row: $d_{cc} = 1.05 \mu m$.  Scale bar: 4 $\mu m$. b) Calculated size, and standard deviation, of the particles cores in images as in (a).
}
 \label{Fig.cryst_struct}
\end{figure}

\section*{Conclusions}

The results presented here show that the interfacial tension value of the bare fluid interface has a profound effect on both the single particle conformation, and on the structural features and mechanical properties of 2D assemblies of soft microgels.
The conformation of the adsorbed microgels can be controlled by varying the oil phase, with lower interfacial tension values (for similar polymer solubility) that cause a decreased deformation of the internal crosslinked core. Microgels at the MTBE/water interface are thicker and less stretched out at the interface. The similar height profiles obtained at the toluene/water and hexane/water interface moreover indicate that above a $\gamma$ threshold value ($\gamma \geq 36 mN \cdot m^{-1}$) the microgel is already fully stretched on the interface plane to minimize the energy of the interface. Interestingly, a similar stretching of the outer polymer chains was evidenced for all oils and microgels investigated, implying that the microgels are still highly surface active at these interfaces, and that the outer polymer chains always adsorb on the interface plane to decrease unfavorable fluid-fluid contacts. 

The resulting 2D assemblies show qualitatively similar structures, but different mechanical responses. Microgels at the hexane/water interface are less deformable during interface compression and, consequently, the resulting monolayer are stiffer. At high compression, the polymer composing the coronae remain on the fluid surface, separating the microgels cores, up to the completion of the isostructural phase transition. Conversely, the lower $\gamma$ value at the MTBE/water interface facilitates the compression of the coronae, shifting the onset of the isostructural phase transition to larger relative interparticle distances, \textit{i.e.} normalized by the size of the corresponding isolated microgels at the interface. Afterwards, the compressed coronae readily collapse onto the cores presumably due to desorption of the polymer from the interface.

Changes in the single particles shape at the fluid interface can be exploited to modulate the microstructural properties of the resulting assemblies in light of applications. As evidenced by Fig.~\ref{Fig.cryst_struct}(a), the crystalline structures obtained as a function of $\gamma$ and $\Pi$ show the possibility of independently tuning the core size and the interparticle spacing in the monolayer. While $\Pi$ gives control over the latter quantity, $\gamma$ affects the in- and out-of-plane deformation of the microgels. As a result, smaller (and thicker) particle cores are obtained, at constant interparticle distances, when MTBE is used at the top phase (Fig.~\ref{Fig.cryst_struct}(b)). Consequently, different ratios of size of the core \textit{versus} that of the corona can be reached. Therefore, the choice of the top fluid phase add an orthogonal switch to further control the microgel organization, which is of particular interest for example for patterning applications.

We expect that our findings will stimulate additional investigations on detailed interplay between the specific nature of the two fluids forming the interface, not just in terms of interfacial tension but also on relative solubility differences, to extend the broad range of factors determining the fascinating response of soft particles at fluid interfaces.


\section*{Declaration of Competing Interest}
The authors declare that they have no known competing financial interests or personal relationships that could have appeared to influence the work reported in this paper.

\section*{Author Contribution Statement}
Author contributions are defined based on the CRediT (Contributor Roles Taxonomy) and listed alphabetically. Conceptualization: J.B., P.F., N.N., L.I, J.V. Formal Analysis: J.V. Funding acquisition: P.F., L.I., J.V. Investigation: N.N., J.V. Methodology: L.I., J.V. Project Administration: J.B., P.F., L.I., J.V. Resources: J.V. Supervision: J.B., P.F., L.I. Validation: J.B., J.V. Visualization: L.I., J.V. Writing - original draft: L.I., J.V. Writing - review and editing: J.B., P.F., N.N., L.I, J.V.

\section*{Acknowledgements}
J.V. acknowledges funding from the European Union’s Horizon 2020 research and innovation programme under the Marie Skłodowska Curie grant agreement 888076.


\bibliography{bibliography}

\newpage

\section*{Graphical Abstract}

\begin{figure*}[h!]
 \centering
 \includegraphics[scale=1]{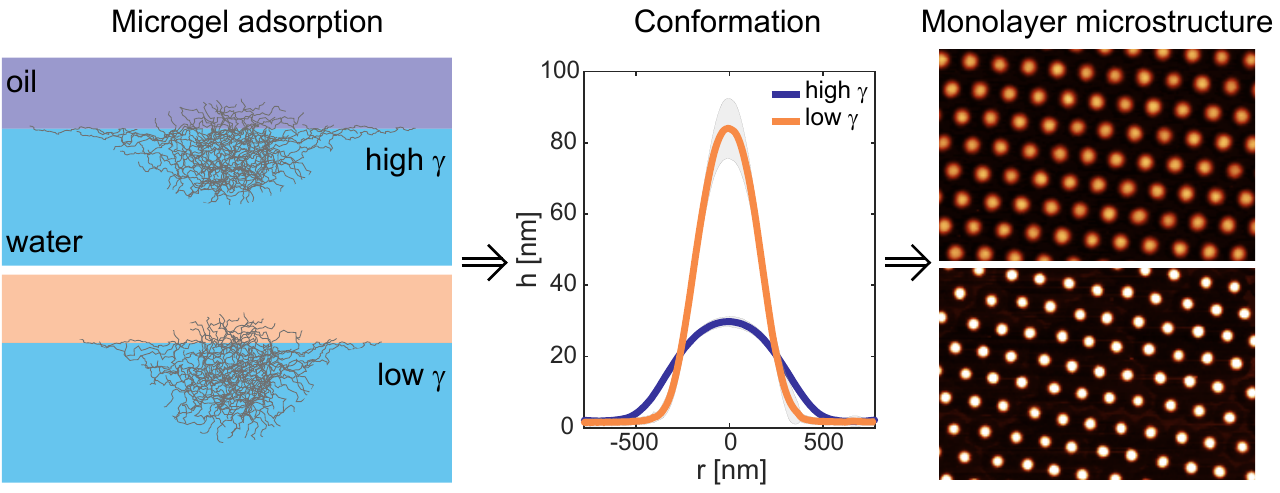}
 \label{Fig.ToC}
\end{figure*}


\newpage


\begin{center}
\vspace*{\fill}
\LARGE
\textbf{Supplementary Material}
\vspace*{\fill}
\end{center}

\setcounter{figure}{0}
\renewcommand{\thefigure}{S\arabic{figure}}
\renewcommand{\thetable}{S\arabic{table}}

\newpage

\section{Supplementary Figures}

\begin{table}[h]
\small
  \caption{\ Microgels hydrodynamic diameters ($D_h$) in aqueous solution} 
  \label{tbl:example1}
  \begin{tabular*}{1\textwidth}{@{\extracolsep{\fill}}llll}
    \hline
    Microgel & $D_h$ at 22$^{\circ}$C [nm] & $D_h$ at 40$^{\circ}$C [nm] & Swelling ratio \\
    \hline
    Soft (1 mol \% BIS) & 786 $\pm$ 13 & 319 $\pm$ 1 & 2.46 $\pm$ 0.02 \\
    Stiff (5 mol \% BIS) & 918 $\pm$ 17 & 499 $\pm$ 2 & 1.84 $\pm$ 0.02 \\
    \hline
  \end{tabular*}
\end{table}

\bigskip

\begin{figure}[h]
\centering
\includegraphics[scale=1]{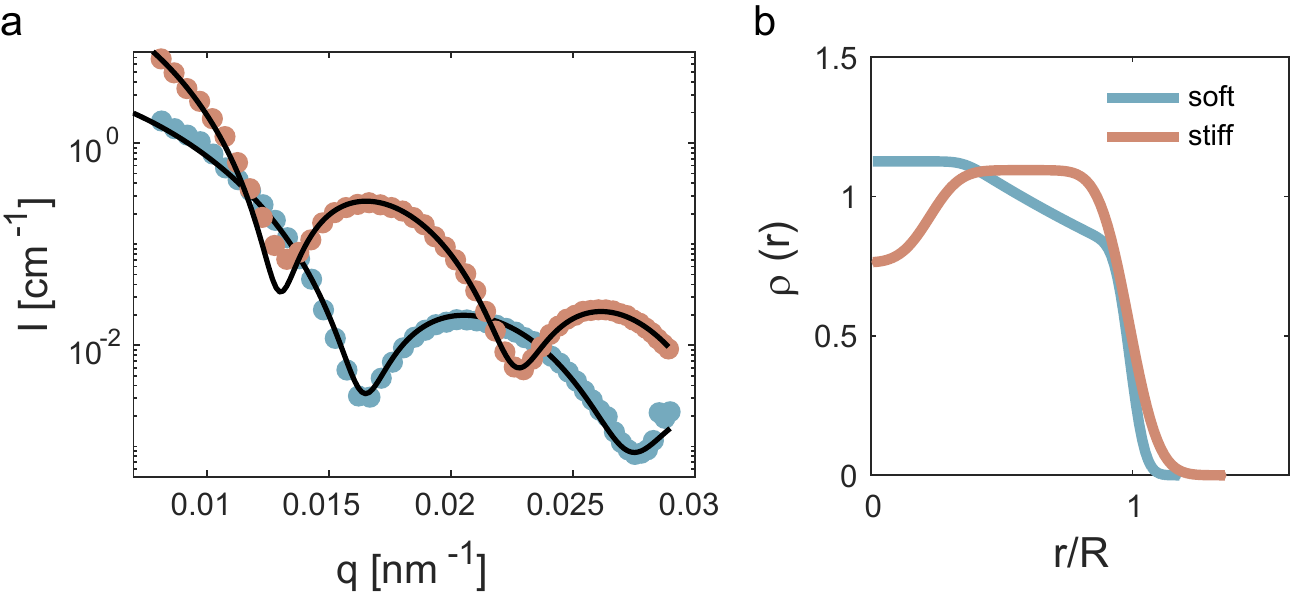}
\caption{\small \textbf{Microgels form factors and polymer density profiles.} a) Experimental form factors obtained from SLS experiments at 25$^{\circ}$C for the soft (1 mol \% BIS) and stiff (5 mol \% BIS) microgels. Black lines are fits (see Methods). b) Microgel radial density profiles ($\rho(r)$) plotted as a function of a normalized radial coordinate $r/R$, where $R$ is the particle radius, as extracted from the fitting procedure of the experimental form factors, as explained in the Methods section.}
\label{fig:SLS}
\end{figure}

\begin{figure}[!htb]
\centering
\includegraphics[scale=1]{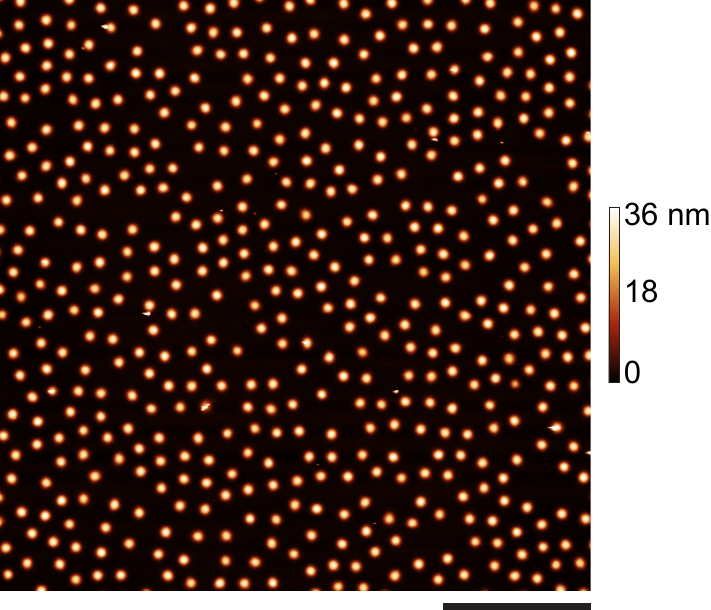}
\caption{\small \textbf{Monolayer structure at low particle concentration.} AFM height image of the monolayer transferred from the hexane/water interface onto a silicon wafer, at surface pressure $\Pi \simeq 1 \cdot 10^{-3} N/m$ and average $d_{cc} \simeq 1.9 \mu m$. Scale bar: 10 $\mu m$.
}
\label{fig:substrates}
\end{figure}


\begin{figure}[!htb]
\centering
\includegraphics[scale=1]{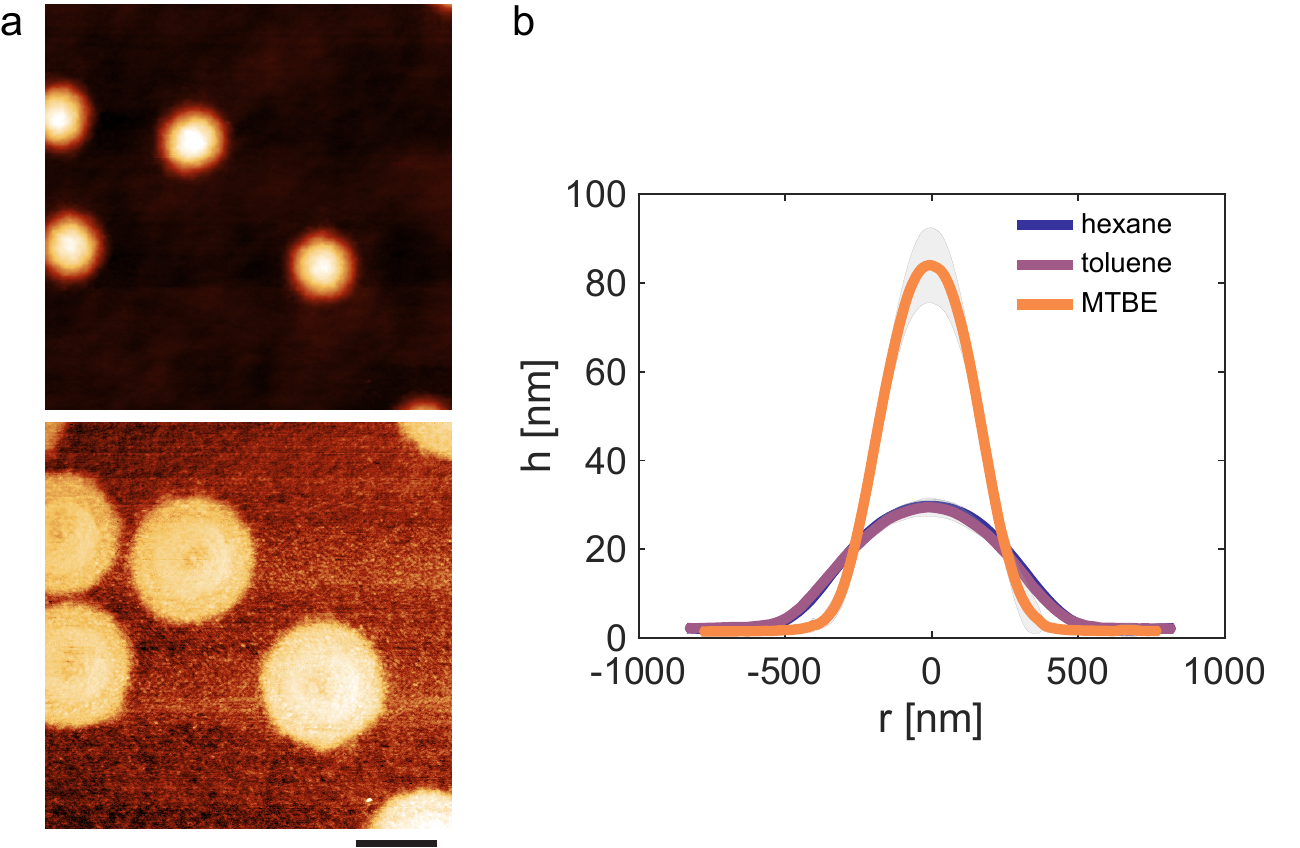}
\caption{\small \textbf{AFM profiles of individual microgels adsorbed at the toluene/water interface, and comparison among different oil phases} a) Representative AFM height (top) and phase (bottom) images of microgels transferred from the toluene/water interface. Scale bar: 1 $\mu$m. b) Experimental height profiles of microgels transferred from the toluene/water interface (purple curve). For comparison, also profiles from the hexane/water (blue curve) or MTBE/water (orange curve) interface are plotted. The shaded regions correspond to the standard deviations of the height profiles calculated on around 20 particles.}
\label{fig:toluene}
\end{figure}

\begin{figure}[!htb]
\centering
\includegraphics[scale=0.9]{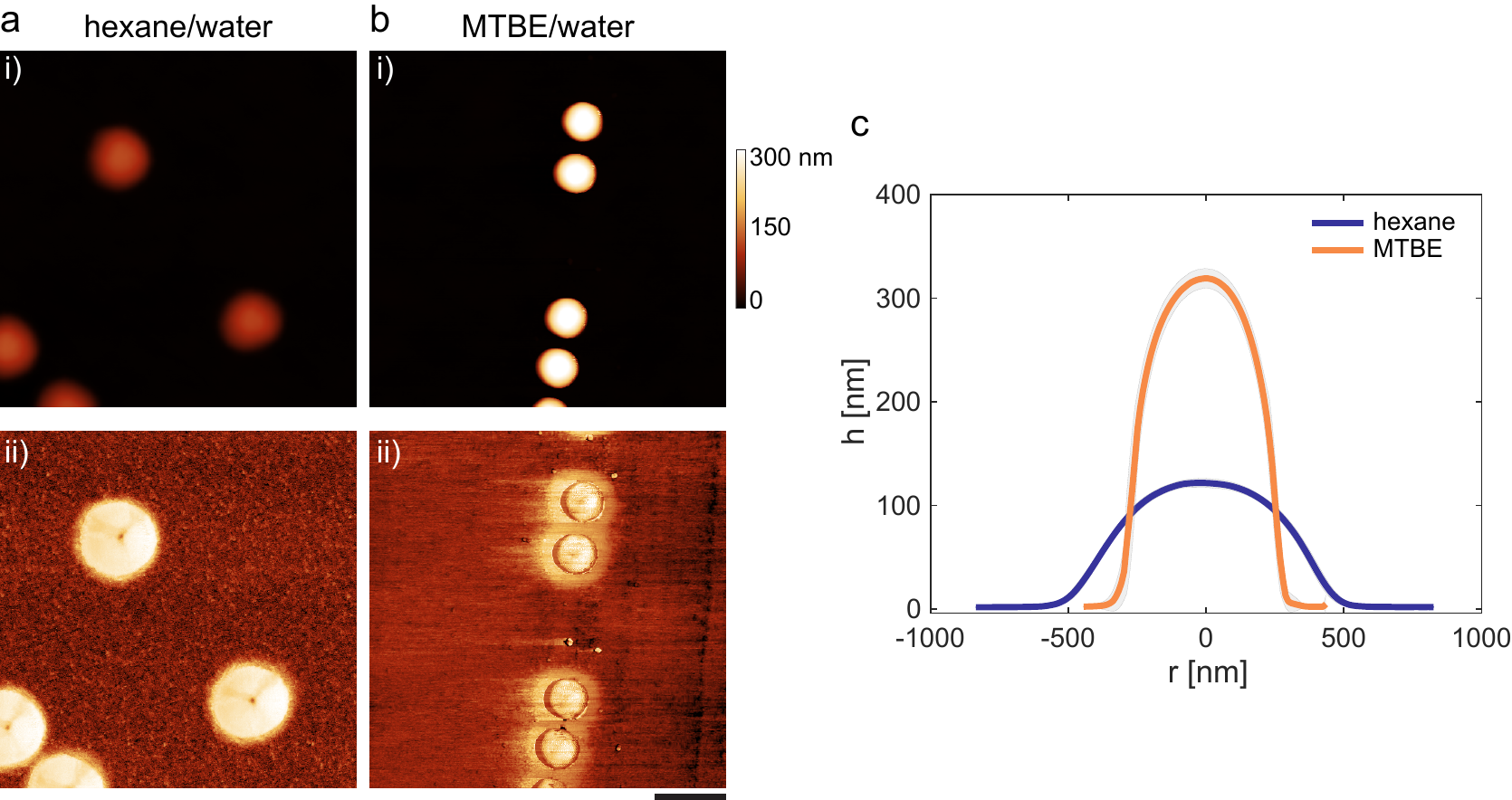}
\caption{\small \textbf{Conformation of stiffer microgels (5 mol \% BIS) adsorbed at the oil/water interface and transferred onto a silicon wafer.} a-b) Representative AFM height (i) and phase (ii) images of microgels transferred from the hexane/water (a, left column) or MTBE/water (b, right column) interface. Scale bar: 1 $\mu$m. c) Experimental height profiles of microgels transferred from the hexane/water (blue curve) or MTBE/water (orange curve) interface. The shaded regions correspond to the standard deviations of the height profiles calculated on around 20 particles.}
\label{fig:ug-23}
\end{figure}

\begin{figure}[!htb]
\centering
\includegraphics[scale=1.1]{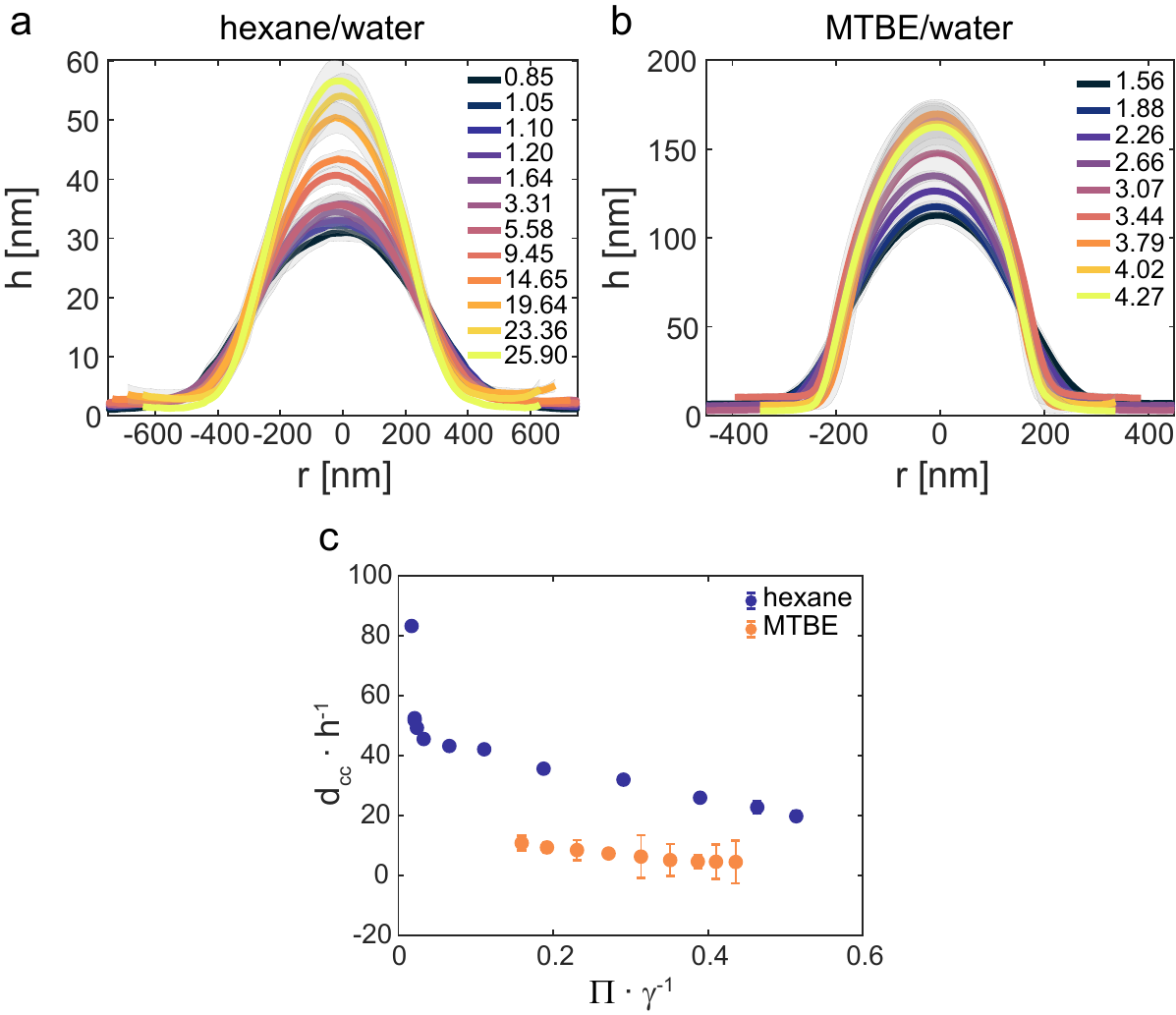}
\caption{\small \textbf{Microgel height profiles as a function of the surface pressure.} Single-microgel height profiles at increasing $\Pi$ (in $mN \cdot m^{-1}$), before the isostructural phase transition. Monolayers are transferred from the hexane/water (a) or MTBE/water interface (b), respectively. The shaded regions correspond to the standard deviations of the height profiles calculated on around 10 particles.}
\label{fig:single-ug_comp}
\end{figure}

\begin{figure}[!htb] 
\centering
\includegraphics[scale=0.7]{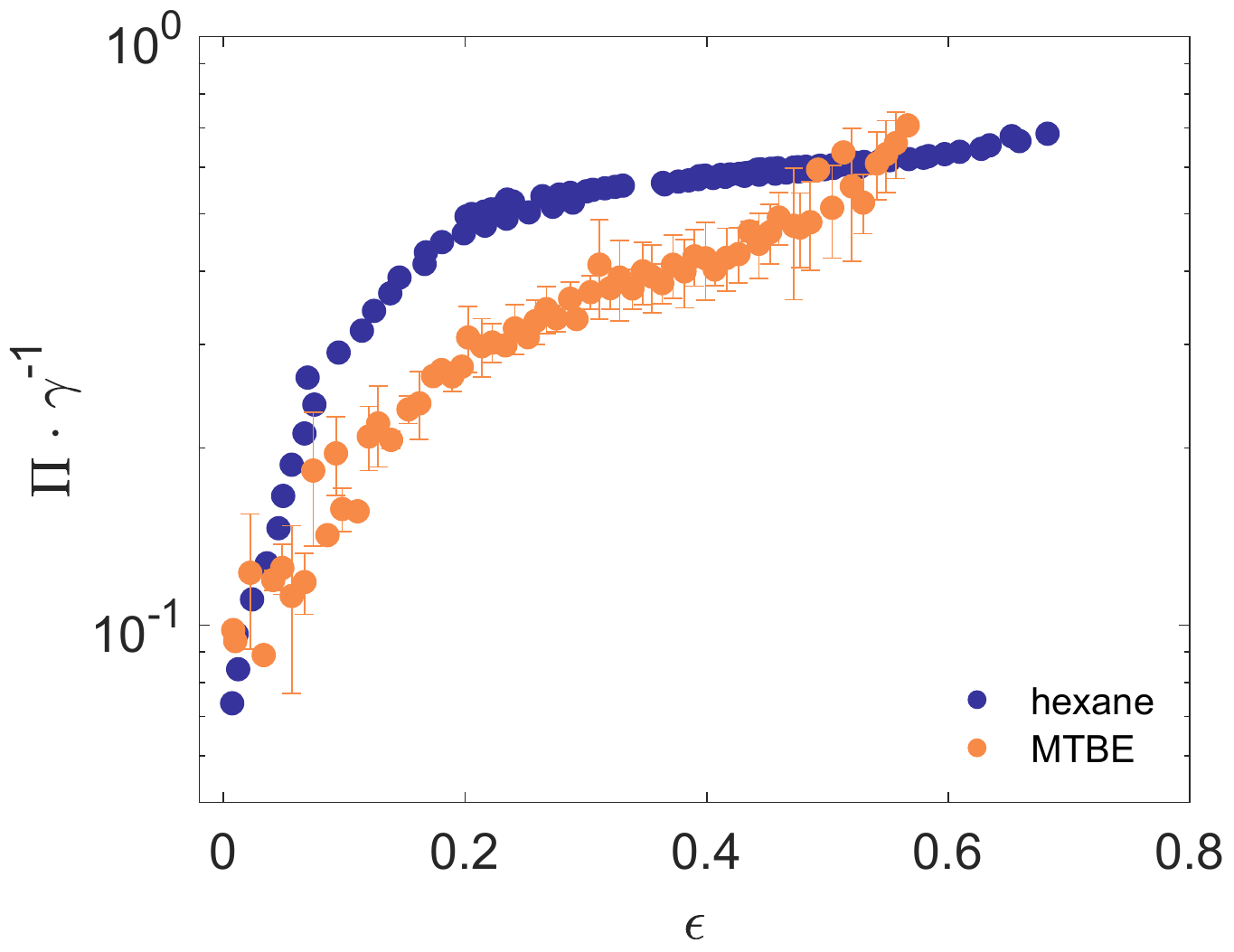}
\caption{\small \textbf{Average compressive strain in monolayers at the hexane/water and MTBE/water interface.} Plot of the normalized surface pressure \textit{versus} the average compressive strain in the monolayers ($\epsilon$), calculated as $\epsilon = (\sigma - d_{cc}) / \sigma$.
}
\label{fig:strain}
\end{figure}

\end{document}